\documentclass[a4paper,twocolumn]{article}

\usepackage[english]{babel}
\usepackage[top=2cm,bottom=2cm,left=2cm,right=2cm,marginparwidth=2cm]{geometry}

\usepackage{amsmath}
\usepackage{graphicx}
\usepackage[colorlinks=true, allcolors=blue]{hyperref}
\usepackage{easyReview}
\usepackage{csquotes}
\usepackage{authblk}
\usepackage{siunitx}
\usepackage{xcolor}

\usepackage[backend=biber, sorting=ynt]{biblatex}

\title{
Direct Evidence of Apex–Hypha Interactions During Vegetative Growth of Fungal Thallus via Comprehensive Network and Trajectory Extraction
}

\author[1]{Thibault Chassereau}
\author[2]{Florence Chapeland-Leclerc}
\author[1,*]{\'Eric Herbert}
\affil[1]{Université Paris Cité, CNRS, UMR 8236-LIED, F-75013 Paris, France}
\affil[2]{Université Paris Cité, CNRS, UMR 8038-CiTCoM, F-75006 Paris, France}
\affil[*]{Corresponding author : eric.herbert@u-paris.fr}

\renewcommand{\vec}{\boldsymbol} 
\newcommand{\mean}[1]{\left<#1\right>}
\begin{document}

\maketitle

\begin{abstract}
The mycelium of a filamentous fungus is a growing, branching network of numerous entangled hyphae exhibiting polarised apical growth. Expansion occurs during the vegetative phase from a single ascospore, driven by the need to explore and occupy surrounding space—limiting competitors, enhancing nutrient uptake, and promoting spore dispersal.
Radial, rapid, and rectilinear growth combined with frequent branching appears adaptive. However, passive growth without interactions or feedback may produce suboptimal networks, as neither local density nor potential connectivity is considered. Reorientations of the apex near existing hyphae suggest apex–hypha feedback. Yet, the diversity of behaviours, spontaneous fluctuations, and limited apical trajectories studied leave open the question of active regulation.
To investigate possible apex–hypha interactions, we analyse a dataset of \textit{Podospora anserina} thallus growth by reconstructing all apical trajectories post-branching and fitting them with a classical Langevin model that incorporates potential interactions.
Comparing isolated and non-isolated  hyphae trajectories allows to identify a clear signature of interaction composed of abrupt deceleration and reorientation. This work opens the path towards a systematic exploration of hyphal interactions.
\end{abstract}

\section{Introduction}

 The growth of spatial branching networks is currently the focus of intense scientific investigation, largely driven by the emergence of new experimental setups that allow for time-lapse imaging of developmental dynamics such as root system~\cite{fernandez_high-throughput_2022} or blood network~\cite{zukowski_breakthrough-induced_2024}. This allows for the development of reconstruction methods aimed at interpreting networks lacking growth time-lapse recordings, \textit{eg} urban road network~\cite{el_gouj_urban_2022} {or} very slowly growing networks such as gorgonians~\cite{douady_work_2020}. Among such systems, filamentous fungi~\cite{oyarte_galvez_travelling-wave_2025, dikec_hyphal_2020, vidal-diez_de_ulzurrun_modelling_2017, du_morphological_2016} are of particular interest, as they exhibit the key characteristics of branching networks: continuous growth, structural persistence, branching behavior, and the formation of anastomoses. A central question in understanding the architecture of these networks concerns their capacity for self-interaction that is, the ability to adapt growth patterns in response to the proximity of other parts of the same network. This may involve, for instance, self-avoidance to optimize biomass distribution, or self-attraction to enhance internal connectivity.
 
The thallus of filamentous fungi, also known as mycelium, is a typical case for growing and branching networks. 
It consists of a collection of static interconnected hyphae, long cylinders from \qty{2}{} to \qty{20}{\um} in diameter, whose growth is polarized, \textit{ie} the extension is localized at the tip (or apex)~\cite{sanati_nezhad_cellular_2013}.
Each apex eventually branches, giving rise to new apexes~\cite{boswell_modelling_2012}.
The ensemble of hyphae then forms a complex, dynamic living network, whose evolution is determined by the individual behavior of each hypha and by the arrangement of these filaments among themselves ~\cite{islam_morphology_2017}.
Then, the fungal network allows an efficient spatial exploration and exploitation of the nutritive resources and provides an effective response to external constraints disturbing their environmental exploration, which largely explains why these organisms occupy a much broader spectrum of habitats~\cite{harris_branching_2008, brand_mechanisms_2009, fricker_mycelium_2017}. 

It has been reported that Spitzenkörper (SPK) localization~\cite{wright_optical_2007} drives the orientation of apical extension. 
The SPK spontaneously randomly moves at the vicinity of the apex, which allows for passive (\textit{ie} non-retroactive) exploration of space.
A passive network has the advantage of minimizing the feedback required for its growth but lacks the ability to adapt to the specific environmental constraints especially in highly competitive territory where colonization and competition are crucial~\cite{boddy_fungal_2016} or in the case of invasive growth~\cite{fukuda_trade-off_2021}.
Among the possible feedback mechanisms, self/non-self recognition plays a particularly important role. 
{In filamentous fungi, this process is formalized as vegetative incompatibility, where contact between genetically distinct individuals triggers programmed cell death and prevents stable hyphal fusion, thereby maintaining functional individuality within the mycelial network ~\cite{paoletti_vegetative_2016}.}
{It could allow for the optimization of network density—and consequently biomass production—in complex systems where density may otherwise quickly become dramatically uneven. }
{By restricting anastomosis between incompatible individuals, these systems limit cytoplasmic mixing and avoid the formation of poorly regulated, over-connected networks, which contributes to more efficient resource allocation at the scale of the mycelium.}
Additionally, it contributes to spatial occupation that limits access to competitors, and it plays a role in sexual reproduction. 
{In fungi, self/non-self recognition also functions as a defense mechanism analogous to innate immunity, as it restricts the horizontal transmission of deleterious elements such as mycoviruses across the network ~\cite{wu_virus-mediated_2017}. Experimental disruption of these recognition systems leads to increased fusion between incompatible strains and facilitates the spread of such elements, highlighting their key role in maintaining territorial integrity and genetic stability.}
Feedback on growth associated with self-recognition is autotropism.

The phenomenon of autotropism between hyphae plays a key role in the overall organization of the mycelium. Negative autotropism describes the repulsive influence of hyphae on each other, whereas positive autotropism, by which hyphae attract each other, can lead to fusion, or anastomosis, particularly in ascomycetes and basidiomycetes~\cite{simonin_physiological_2012}.
Autotropism has been reported as intermittent, sometimes positive, sometimes negative, and often intertwined with other mechanisms—most notably spontaneous hyphal curvature—making its direct identification inherently challenging. To our knowledge, studies on autotropism have relied on manually collected, isolated events that are prone to selection bias, and have never been conducted within the context of a complete hyphal network, primarily due to experimental limitations.

In this study, we examine the growth of the filamentous fungus \textit{P. anserina} by tracking all of its apexes within a previously reconstructed network.
To investigate the role of spatial context, we divide the apexes into two groups according to their proximity to the existing mycelial network: those considered to grow in the absence of interaction, corresponding to development in a homogeneous and isotropic medium, and those likely to be influenced by interactions with the surrounding structures.
We then analyse the trajectories of these apexes through the framework of Langevin’s drift–diffusion equation, applying the model separately to each group in order to compare their dynamical behaviours.

\begin{figure*}
    \centering
    \includegraphics[width=.9\linewidth]{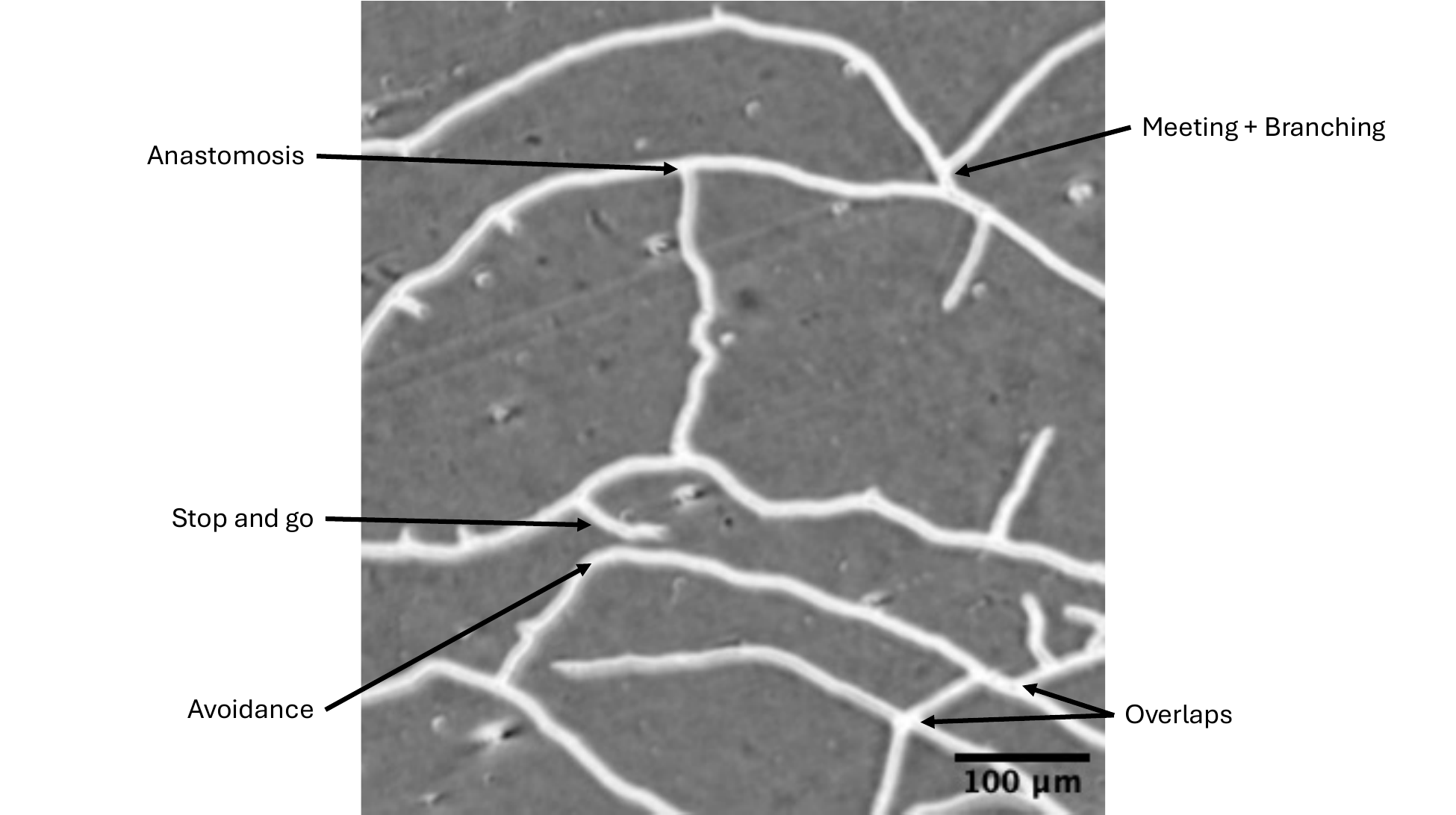}
    \caption{Examples of possible interactions within a small portion of the mycelium of \textit{P. anserina}. From top to bottom and left to right, the reader can observe a probable anastomosis, a meeting followed by a branching event, a stop-and-go, an avoidance and two successive overlaps. The supplementary materials offer the opportunity to view the movie of the growth of this section of the mycelium.
    }
    \label{fig:ExampleInteractions}
\end{figure*}

\section{Materials and Methods}

\subsection{\textit{Podospora anserina}}\label{sec:annex_Panserina}
\textit{P. anserina} is a saprophytic ascomycete filamentous fungus that makes an excellent laboratory model, in particular because its life cycle is short (around 7 days) and perfectly controlled in vitro on a standard culture medium~\cite{silar_podospora_2020}. 
In order to analyse the growth dynamics of \textit{P. anserina} under controlled conditions, a two-dimensional image acquisition and processing system were developed in our laboratory~\cite{dikec_hyphal_2020}, see figure\,\ref{fig:ExampleInteractions}. In particular, it has been used to characterize the behavior of a fungal thallus under various constraints~\cite{ledoux_characterization_2024} ; and to propose a complete identification of a growing and branching network’s spatio-temporal structures thanks to a general method for global network reconstruction~\cite{chassereau_full_2025}.
The reconstruction of the thallus as a spatio-temporal graph enables the individual tracking of all apexes and the reconstruction of each growth trajectory. This yields a complete collection of apical trajectories, that take into account its environment, particularly the location of the nearest hyphae at any given moment during growth.
The dataset is divided into two subsets: the first corresponds to growth in the absence of nearby hyphae, allowing characterisation of velocity in a homogeneous and isotropic environment; the second includes potential interactions with neighbouring hyphae,  see figure\,\ref{fig:tortuosity}-A.
These collections are subsequently compared to a stochastic growth model that incorporates both passive extension dynamics and active orientation mechanisms, with autotropism as a central component.
The work presented here is based on the growth of five \textit{P. anserina} mycelia under standard conditions {(Labeled from A to E in the following and in particular in the table \ref{tab:tab_results})}, as previously described in~\cite{ledoux_caracterisation_2023}. 
{We have previously shown that the conditions under which these networks are cultivated lead to unbiased growth with no fundamental macroscopic differences between these five networks. Furthermore, although the number of networks may seem small, the trajectories we are looking at here are very numerous, even for each network taken individually. Finally, the aim of this article is to show that there is a difference in growth between free branches and obstructed branches by providing a simple criterion for distinguishing between the two cases a priori.}

\subsection{Modelling}\label{sec:modelisation}
We are interested in modelling the growth of hyphae forming the thallus of a filamentous fungus on a homogeneous and isotropic culture medium. The hyphae grow at their apex, whose displacement forms a trajectory.  {We consider growth to occur in two dimensions. The third spatial dimension is not considered because a sheet of cellophane between the thallus and the culture medium prevents the hyphae from penetrating the medium. The resulting thallus is quasi-flat. However, hyphae overlap is still possible and can be considered as a form of interaction between hyphae.}
Experimentally, we observe apex trajectory over time  to be largely rectilinear, with noise of biological origin, linked to the position of the SPK~\cite{reynaga-pena_analysis_1997, riquelme_spitzenkorper_2014}. 
{No true inertial phenomena are involved — the kinetic energy of the apex plays no role in filament elongation. However, apical growth does exhibit a characteristic relaxation timescale a half an hour, visible in the initial acceleration phase~\cite{ledoux_prediction_2023}, which suggests an effective-inertia term. We build on stationary and Gaussian biological noise assumptions, allowing us to make use of the Langevin equation as a phenomenological model for apical trajectories: }
\begin{equation}
  \frac{\textrm{d}\vec{v}}{\textrm{d}t} = \vec{F_g} - \gamma \, \vec{v} + \vec{\eta}+\vec{I}
  \label{eq:model_vect}
\end{equation}
{with $\gamma \, \vec{v}$ a drag force  and $\vec{\eta}$ a Gaussian random forces of biological origin  and $\vec{v}$ the growth elongation vector, \textit{ie} the time derivative of the apex position.}
$\vec{F_g}$ correspond to the effective growth force. 
Its origin can be multiple (turgor pressure, osmotic pressure, etc). 
We make the hypothesis that its norm is constant and that it is always aligned with the growth elongation vector $\vec{v}$ as the tip of the hypha is softer than the rest of the cell~\cite{fukuda_trade-off_2021}.
The term $\gamma$ {represents} the resistance {of} the medium and the limited capacity of the fungus to produce and transport to the SPK the material needed for the growth of the hyphae.
The direction of hypha growth depends on the localisation of the SPK, $\vec{\eta}$  represents the variable position of this vesicles cluster. 
We make the hypothesis that this relative position follows a 2D isotropic Gaussian noise without correlation in time.
Moreover, the hyphae grows in a complex environment and can interact with each other. 
The term $\vec{I}$ is the result of all these contribution and should, at least, be a function of time and spatial localization. 
As shown in Figure~\ref{fig:ExampleInteractions} (and in the associated movie), these interactions can be attractive as well as negative and occur simultaneously in a complex and changing environment, making it difficult to extract a more precise mathematical expression for $\vec{I}$. 

{We aim to distinguish the effect of interactions in the kinematics of growth. The Langevin framework assumes that all stochastic contributions are Gaussian. However, the interaction term $I$ is not expected to be Gaussian in general, as it reflects discrete, spatially localized events whose statistics are inherently non-Gaussian. As a consequence, the Langevin model cannot reproduce obstructed trajectories by construction, and we do not attempt to fit them within this framework. 
Instead, we will use it as a rigorous baseline: interactions are identified as statistically significant, non-Gaussian deviations from the free-trajectory signature, without requiring an explicit expression for $I$.}

{In order to separate the variations in the elongation vector $\vec{v}$ due to variations in its norm $v$ from those due to variations in its orientation $\theta$, we introduce the moving Frenet reference frame, whose origin is at each moment the position of the apex and is formed by the unit vectors $\vec{u_v}$, tangent to the elongation vector $\vec{v}$, and $\vec{u_\theta}$, perpendicular to $\vec{u_v}$. 
This frame change over time and in particular we have that $\vec{v} = v \vec{u_v}$ and $\frac{\textrm{d}}{\textrm{d}t}\vec{u_v} = \frac{\textrm{d}\theta}{\textrm{d}t} \vec{u_\theta}$. 
Projecting equation~\ref{eq:model_vect} onto $\vec{u_v}$ and $\vec{u_\theta}$, it follows that:}

\begin{equation}
  \left\{ 
  \begin{array}{ccl}
    \frac{\textrm{d}\vec{v}}{\textrm{d}t}
        & = &\gamma\,(v_\infty-v)+\eta_v\\
    v\,\frac{\textrm{d}\theta}{\textrm{d}t}
        & = &\eta_\theta 
  \end{array} 
  \right.
  \label{eq:model_system}
\end{equation}

With $v_\infty$ the apical asymptotic growth speed and $\vec{\eta}=\eta_v \, \vec{u_v}+\eta_\theta \, \vec{u_\theta}$.
Under the assumption that $\vec{\eta}$ is isotropic and follows a normal distribution centred at $0$, with variance $\sigma^2$ and without initial growth speed, it follows that $\mean{\theta}=0$ and $\mean{v}$ time series can be written as:
\begin{equation}
  \mean{v}(t) = v_\infty \left( 1 - \exp{\left(-\gamma \, t \right)}\right)
  \label{eq:v_mean}
\end{equation}
with $\mean{\cdot}$ the ensemble average of $(\cdot)$.
From the equations~\ref{eq:model_system}, it is possible to obtain an estimation for $v_\infty$ and $\gamma$ given an empirical distribution of theses growth vectors.
First, $v_\infty$ is derived immediately:
\begin{equation}
  \begin{array}{ccl}
    v_\infty& = &\mean{v}+\frac{1}{\gamma}\mean{\frac{\textrm{d}\vec{v}}{\textrm{d}t}}
  \end{array} 
  \label{eq:gamma-v_infty}
\end{equation}
Second, by equating the variance of $\eta_v$ and $\eta_\theta$ in equations~\ref{eq:model_system}, and bearing in mind that $v$ and $\dot{v}$ are independent, \textit{ie} the covariance vanishes, we can obtain the following expressions for $\gamma$:
\begin{equation}
  \begin{array}{ccl}
    \gamma& = & \sqrt{\frac{\mean{v^2 \, \dot{\theta}^2} - \text{Var}(\dot{v})}{\text{Var}(v)}}
  \end{array} 
  \label{eq:gamma}
\end{equation}
with $\textrm{Var}(\cdot)$ the variance of $(\cdot)$.

Given the time series of $v$ and $\theta$ of non-interacting apex trajectories, the estimations of $\gamma$, $v_\infty$ and $\sigma = \sqrt{v^2 \, \dot{\theta}^2}$, and following equations\,\ref{eq:model_system} we are then allowed to estimate $\eta_v$ and $\eta_\theta$ as respectively $ \frac{\textrm{d}\vec{v}}{\textrm{d}t} - \gamma\,(v_\infty-v)$ and $v \, \frac{\textrm{d}\theta}{\textrm{d}t}$.

\subsection{Interaction criterion}\label{sec:detection-area}

In this section, we introduce an interaction criterion that allows us to divide the collection of trajectories into two subsets. In the first subset, no interaction is assumed, whereas in the second, at least one interaction is considered possible.

Branching processes have been reported to occur either apically or laterally, depending on the latency between the passage of the parent hypha and the emergence of the branch. These two types of branching have been shown to differ in their growth velocities, as reported in~\cite{chassereau_full_2025}. During the reconstruction procedure, branches are automatically classified as apical or lateral according to the method described in~\cite{chassereau_full_2025}. Lateral branches typically emerge in the central regions of the thallus, where density is highest. This prevents the construction of a collection of non-interacting lateral branches. Consequently, in this work we restrict our analysis to apical branches only.

\begin{figure*}[!ht]
  \centering
  \includegraphics[width=.9\linewidth]{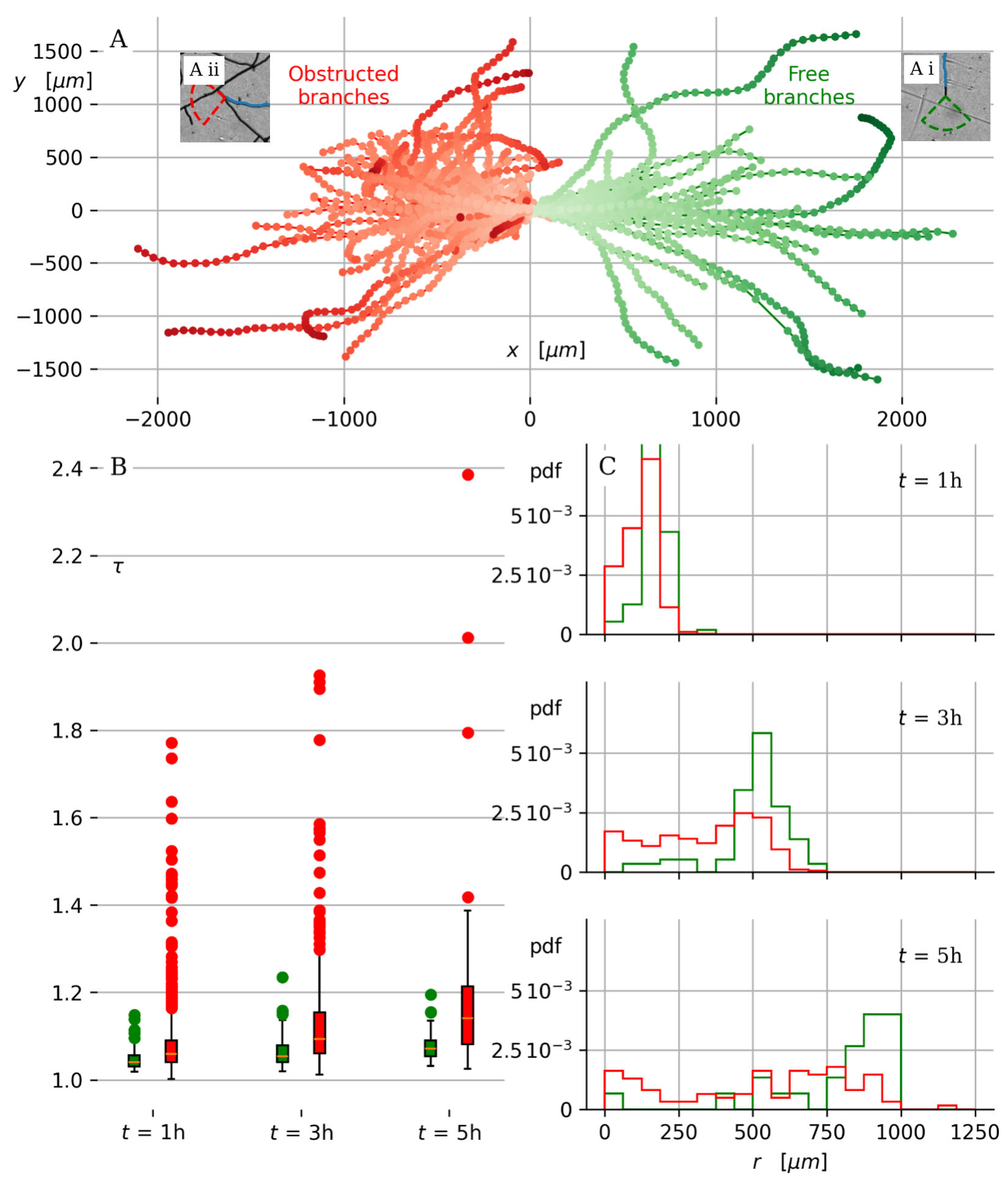}
  \caption{A) Set of aligned trajectory of free branches (initial growth vector aligned to the right) and obstructed branches (initial growth vector aligned to the left).{Insert A-i) and A-ii) are example of respectively free (i) and obstructed (ii) interaction area on top of an apex.}B) distribution of tortuosity $\tau$ of free branches (in green) and obstructed branches (in red) after 1h, 3h and 5h of branch growth. The boxes represent the range of distribution between the first and third quartiles, with an orange line indicating the median. The whiskers show the range extending to the distribution extremes, or to 1.5 times the interquartile range, whichever is closer to the median. The dots correspond to measurements that fall outside this range. C) density function estimate of distance $r$ between starting point and the apex of free branches (in green) and obstructed branches (in red) after 1h, 3h and 5h of branch growth. {Kolmogorov-Smirnov tests have confirmed that the distributions of free and obstructed branches are statistically different at any time and in both cases (part B and C of the figure) (p-value always lower than $3\,10^{-4}$).}}
  \label{fig:tortuosity}
\end{figure*}

The two collections, corresponding respectively to growth with and without interaction, are constructed on the basis of an interaction criterion. An interaction region is defined as a disk sector centered on the apex and oriented along its growth direction. The dimensions of this sector are chosen so as to include all surrounding biomass, excluding the apex itself, in the vicinity of the extending hypha. \\
The angular aperture of the disk sector, $\alpha = \pi/2$, ensures the exclusion of apexes originating from branches of the same hypha. Its radial extent, $r = \qty{180}{\um}$, corresponds to the hyphal elongation over two time steps at asymptotic growth velocity. At each time step, the presence or absence of biomass within this interaction region can be assessed. The collection of free hyphae (without interaction) consists of all apical trajectories for which no biomass is ever detected within the defined region. The complementary set defines the collection of non-free hyphae (with potential interaction).
{Inserts in Figure \ref{fig:tortuosity}-A illustrate a growing hypha with its associated disk sector.} 
Two examples are provided: A-i corresponds to a free hypha, while A-ii represents a non-free hypha. In this and subsequent figures, green is used to denote free hyphae and red to denote non-free hyphae.

It was reported in~\cite{sanchez-orellana_automated_2018} \textit{stop and go} behaviour that could be misinterpreted as an interaction {even though this behaviour does not require the presence of any hyphae other than the one that is growing.} Consequently, branches that are observed to stop at least one time step were removed. In other words, all the apexes coming from an apical branching point, whose growth can be followed long enough to extract at least one acceleration vector, are used.

Note that the spatial and temporal distributions of branch formations are not homogeneous. First, hyphae located at the edge of the thallus are more likely to remain isolated, as they grow in regions of lower density. Second, the number of branches increases exponentially, meaning that the shortest hyphae are also the most numerous.

Enumeration details are resumed in the first five columns of table~\ref{tab:tab_results} for all five thallus.

\section{Results}\label{sec:results}

Let us now turn to the comparison of the modelling with an experimental branching and growing network.
We will build on the images of the 2-D growth of the thallus of saprophytic ascomycete filamentous fungi \textit{P. anserina} obtained in the laboratory and described in \cite{dikec_hyphal_2020,  ledoux_caracterisation_2023} and in section \ref{sec:annex_Panserina},  on homogeneous and isotropic medium. 
The time resolution is \qty{18}{\minute} (see \ref{sec:annex_temporalResolution} for a discussion).
The comparison is based on five experiments described in \cite{ledoux_caracterisation_2023}.
The observed growths are initiated from a single ascospore and show, after a typical duration of \qty{20}{\hour}, a dense network of several metres of hyphae and several thousands of branches. \\
A reconstruction of the network is then performed as described in~\cite{chassereau_full_2025} to obtain the spatio-temporal graph of the entire thallus for the five experiments.
The hyphae are reconstructed to take account of branches, anastomoses and crossings.
For each hyphae of each thallus, the trajectory of the apex is reconstructed as the time series of the positions of the hyphal extremity, see blue lines in {inserts Fig.~\ref{fig:tortuosity}-A}. We only consider the hyphae from an apical branching (see section \ref{sec:detection-area}).

Finally we constructed an interaction criterion based on the presence of a hypha near an apex, based on a circular sector-shaped detection area, see \ref{sec:detection-area} for details. Two-collections of apex trajectories are derived, one for which no interaction (named \textit{free} hereafter) is expected and the second for which they are likely to exist (named \textit{obstructed}). 
{Inserts of Figure~\ref{fig:tortuosity}-A illustrate} the growth interaction area (disk sector), showing both cases, free (green) and obstructed (red). See Table~\ref{tab:tab_results} for corresponding enumeration details of all five-experiments.

\begin{table*}
  \centering
  \begin{tabular}{|c|c|c|c|c|c|c|c|c|c|}
  \hline
    Label & $N_{\textrm{free}}$ & $n_{\textrm{free}}$ & $N_{\textrm{obst}}$ & $n_{\textrm{obst}}$ & $\gamma$ & $v_\infty$ & $\Sigma_\theta$ & $\sigma_\theta$ & $\sigma_v$ \\
     &  &  &  &  & [\unit{\per\hour}] & [\unit{\um\per\hour}] & [\unit{\um\per\hour\squared}] & [\unit{\um\per\hour\squared}]& [\unit{\um\per\hour\squared}]\\\hline
    A & \qty{67}{} & \qty{584}{} & \qty{373}{} & \qty{2279}{} & \qty{1.8(2)}{} & \qty{317(4)}{} & \qty{163(7)}{} & \qty{144(5)}{} & \qty{140(5)}{}\\\hline
    B & \qty{47}{} & \qty{451}{} & \qty{271}{} & \qty{1879}{} & \qty{2.0(2)}{} & \qty{322(3)}{} & \qty{139(8)}{} & \qty{105(6)}{} & \qty{100(5)}{}\\\hline
    C & \qty{69}{} & \qty{685}{} & \qty{631}{} & \qty{3715}{} & \qty{2.2(2)}{} & \qty{322(3)}{} & \qty{150(4)}{} & \qty{133(5)}{}& \qty{119(4)}{}\\\hline
    D & \qty{61}{} & \qty{575}{} & \qty{200}{} & \qty{1395}{} & \qty{1.5(2)}{} & \qty{280(3)}{} & \qty{129(5)}{} & \qty{105(5)}{}& \qty{94(4)}{}\\\hline
    E & \qty{36}{} & \qty{322}{} & \qty{120}{} & \qty{905}{} & \qty{1.5(2)}{} & \qty{291(5)}{} & \qty{126(6)}{} & \qty{101(6)}{} & \qty{87(6)}{}\\\hline\hline
    Mean $\pm$ s.d. & n.r. & n.r. & n.r.& n.r. & \qty{1.8(3)}{} & \qty{307(18)}{} & \qty{141(14)}{} & \qty{118(17)}{} & \qty{108(19)}{}\\\hline
  \end{tabular}
  \caption{Numerical details of the division of the population of apical branches into two sub-populations for each of the 5 networks A,B,C,D and E. $n_{\textrm{free}}$ {the size of} the set of accelerations obtained from $N_{\textrm{free}}$ free apical branches while $n_{\textrm{obst}}$ is the {size of the} set of acceleration extracted from $N_\textrm{obst}$ obstructed apical. The corresponding numerical estimates of $\gamma$, $v_\infty$ and $\Sigma_\theta$ are also given with the bootstrap estimated error. From the estimation of $\Sigma_\theta$ we extract $\sigma_\theta$ by correcting it from the estimated numerical error $\epsilon_\theta$ (see text for more details). $\sigma_v$ is calculated the same way, with the corresponding projection.
  }
  \label{tab:tab_results}
\end{table*}

\subsection{Apical extension trajectories}

We first examine the growth trajectories represented by the two sets of branches. 
Figure~\ref{fig:tortuosity}-A displays these trajectories, aligned according to their initial growth vector $\vec{v}$. 
It is important to remember that the initial time corresponds to the birth of the branch. 
For ease of comparison, we oriented free hyphae to the right, while those that encountered an obstruction at any point during their development are initially oriented to the left.

From this representation, it is apparent that the trajectories of free hyphae are visually straighter than those of obstructed hyphae. 
Free branches tend to maintain their initial growth direction, whereas some non-free hyphae undergo such marked deviations that their trajectories may bend back toward their point of origin. 
This effect can be quantified by computing the tortuosity, $\tau$, defined as the ratio of $d$ the displacement (curvilinear branch length) to $r$ the shortest distance. 
Figure~\ref{fig:tortuosity}-B presents the evolution of the tortuosity distribution at different growth periods (2h, 5h, and 8h, respectively).
Free branches exhibit a consistent behavior across time, with tortuosity values ranging from 1.05 to 1.2. 
This suggests that their growth dynamics remain unaltered, as expected in a homogeneous and isotropic medium. 
By contrast, non-free branches display tortuosity values ranging from 1.05 to above 2, with a distribution that broadens over time. 
This increasing tortuosity reflects substantial trajectory modifications, consistent with the expected signature of interactions. 

Since the initial orientation of growth is common to all branches, we can roughly estimate the diffusion of the apexes around a straight trajectory. 
To this end, we represent the probability density functions (pdfs) of the distances $r$ from the origin of each apex, for growth durations of 1h, 3h, and 5h (see figure~\ref{fig:tortuosity}-C).
It is clear that the distribution of $r$ for free hyphae is consistently narrower than that for obstructed hyphae, and that its width increases over time. 
As a first approximation of the diffusion of apexes around the straight trajectory, we compute the diffusion coefficient $D = \frac{\langle (r - \langle r \rangle )^2 \rangle}{4\,t}$.
{The numerical estimates of $D$ and $\langle r \rangle$ after \qty{1}{\hour}, \qty{3}{\hour}, and \qty{5}{\hour} of growth are shown in table~\ref{tab:diffusion}}
For free hyphae, we obtain $D=\:$\qty{420(95)}{}, \qty{1200(270)}{}, \qty{2400(1000)}{\micro\meter^2\per\hour}, with corresponding mean values $\langle r \rangle =\:$\qty{165(4)}{}, \qty{502(12)}{},  \qty{790(45)}{\micro\meter} for \qty{1}{\hour}, \qty{3}{\hour}, and \qty{5}{\hour}, respectively. 
For obstructed apexes, we find $D =\:$\qty{712(34)}{}, \qty{2612(123)}{},  \qty{4576(424)}{\micro\meter^2\per\hour}, with corresponding $\langle r \rangle =\:$\qty{122(2)}{}, \qty{332(10)}{}, \qty{525(30)}{\micro\meter} for the same growth durations.
The distance $\langle r\rangle$ grows approximately linearly for free hyphae, leading to a typical constant propagation velocity of $\langle r \rangle / t=\:$\qty{160}{\um\per\hour}. 
In contrast, for non-free hyphae, the propagation velocity is lower and decreases from \qty{120}{\um\per\hour} to \qty{105}{\um\per\hour}.
It is important to note that this speed differs from the growth speed of hyphae, as it is based solely on the radial distance between the apex and the initial branching point, and not on the total length of the hyphae.
As expected for an advected process, the diffusion coefficient $D$ is not constant but exhibits temporal growth. 
Remarkably, despite the smaller mean displacement from the origin in the non-free case, $D$ exceeds that of the free case by a factor of two, reflecting a substantially enhanced lateral spreading. 

\begin{table*}[]
    \centering
    \begin{tabular}{|c||c|c|c|}
    \hline
         & \qty{1}{\hour} & \qty{3}{\hour} & \qty{5}{\hour} \\\hline\hline
        $\langle r \rangle$ free branches & \qty{165(4)}{\micro\meter} & \qty{502(12)}{\micro\meter} &  \qty{790(45)}{\micro\meter}\\\hline
        $\langle r \rangle$ obstructed branches & \qty{122(2)}{\micro\meter} & \qty{332(10)}{\micro\meter} & \qty{525(30)}{\micro\meter} \\\hline\hline
        $D$ free branches & \qty{420(95)}{\micro\meter^2\per\hour} & \qty{1200(270)}{\micro\meter^2\per\hour} & \qty{2400(1000)}{\micro\meter^2\per\hour}\\\hline
        $D$ obstructed branches & \qty{712(34)}{\micro\meter^2\per\hour} & \qty{2612(123)}{\micro\meter^2\per\hour} &  \qty{4576(424)}{\micro\meter^2\per\hour}\\\hline
    \end{tabular}
    \caption{{Numerical estimation of mean radial distribution $\langle r\rangle$ and effective diffusion coefficient $D = \frac{\langle (r - \langle r \rangle )^2 \rangle}{4\,t}$ for free and obstructed branches after $t = \:$ \qty{1}{\hour}, \qty{3}{\hour} and \qty{5}{\hour} of growth.}}
    \label{tab:diffusion}
\end{table*}

\subsection{Free apical trajectories}
\begin{figure*}[!ht]
  \centering
  \includegraphics[width=1\linewidth]{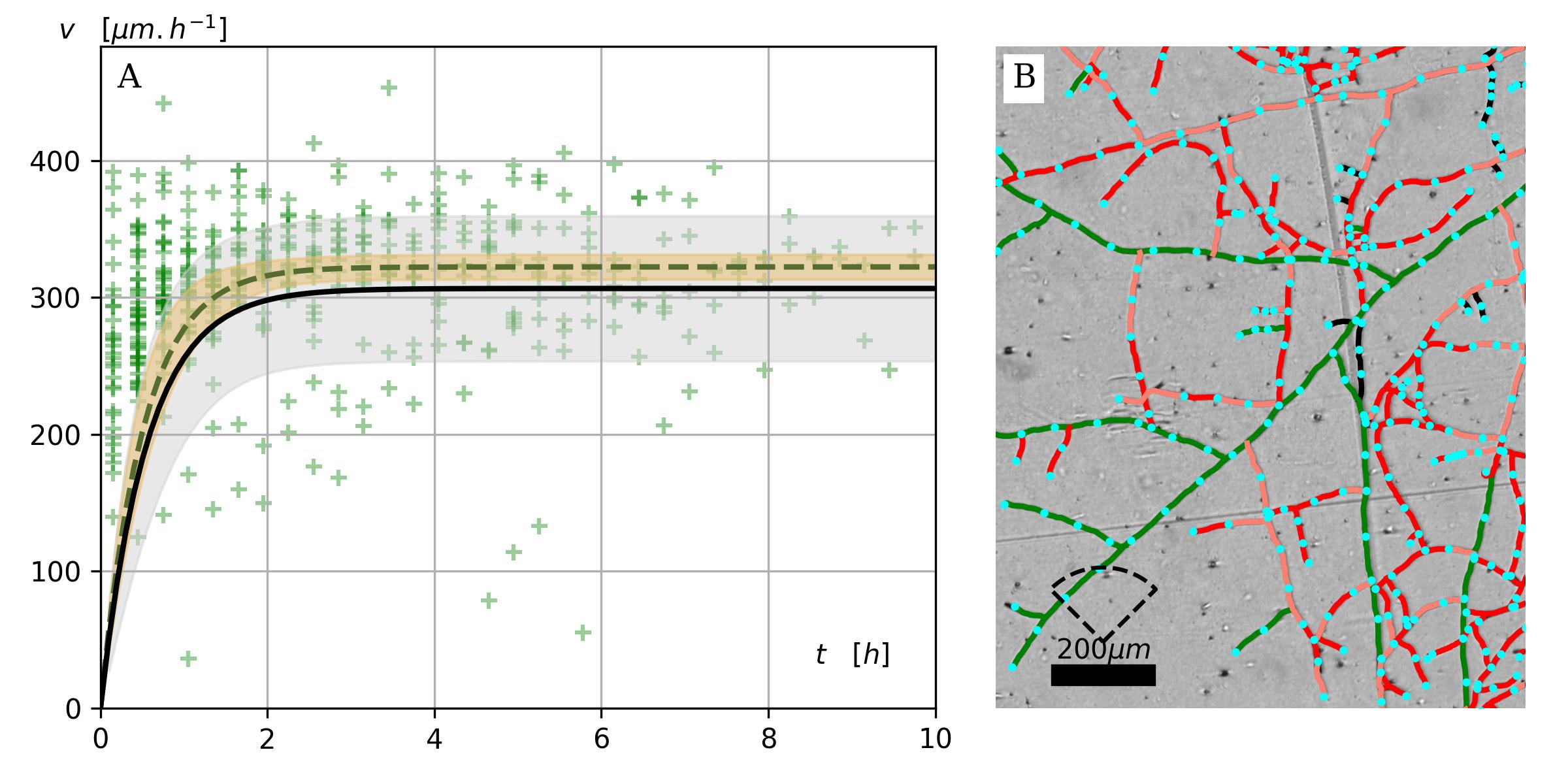}
  \caption{A) scatter plot of elongation speed as a function of time since the start of the growth of unobstructed apical branches for the network B, estimated to be representative of the five networks. Green dashed line correspond to the model evaluated on this network with yellow shade indicating three standard deviation estimated via bootstrap. Black line show the mean model over the five networks with the gray area showing three standard deviation. {B) example of reconstruction of the network C showing, in black, the lateral branches, in green, the free apical branches and in red, the obstructed apical branches (pink portions indicates growth under a free interaction area while darker red indicate growth occurring under obstructed interaction area). Cyan dots represent successive apexes positions. Scale bar: \qty{200}{\um}.}}
  \label{fig:elongation_analysis}
\end{figure*}

Let us now switch to the comparison with the model. From $v$ and $\theta$ time series we derived $\gamma$ (eq.~\ref{eq:gamma}), $v_\infty$ (eq.~\ref{eq:gamma-v_infty}) and $\sigma$. Results for the five networks are summarized in table \ref{tab:tab_results}.
The displacement velocity, $v_\infty\approx \qty{300}{\um\per\hour}$, value is compatible with the ones reported in the literature for \textit{P. anserina}~\cite{silar_podospora_2020} and with estimates previously obtained from the same thalli from manual~\cite{ledoux_prediction_2023} and automatic~\cite{chassereau_full_2025} extractions. The estimation of the acceleration term $\gamma$ is found in agreement with~\cite{chassereau_full_2025}. 

In a first step, figure~\ref{fig:elongation_analysis}-{A} shows the superposition of the experimental velocities with the equation \ref{eq:v_mean} parameterised with the values of $\gamma$ and $v_\infty$ obtained previously.

Except the initial short acceleration phase (approximately 1~hour), hyphae grow at a constant steady state speed as simplified in numerous previous models~\cite{ carver_lattice-free_2008, vidal-diez_de_ulzurrun_modelling_2017}.
In the permanent regime, the fluctuations of speed observed notably in~\cite{sanchez-orellana_automated_2018} at higher temporal resolution, with stops and slowdowns of growth of the mother hyphae at the time of a connection are averaged. 

{Figure~\ref{fig:elongation_analysis}-B illustrates the extracted growth dynamics of the network C.
Each apex location is marked, with spatial separation reflecting growth velocity. 
Notably, lateral hyphae exhibit significantly slower growth compared to apical hyphae. 
They are also shorter and located in regions of higher density.
Focusing on apical hyphae—the primary subject of this discussion—free hyphae demonstrate the predominantly rectilinear behavior described in figures\,\ref{fig:elongation_analysis}-A and \ref{fig:tortuosity}-A, characterized by regularly spaced apex distribution. 
Reorientation events in these hyphae occur over long wavelengths.
When examining the obstructed fraction of obstructed hyphae, a diverse range of behaviors emerges, complicating classification. 
Some trajectories are clearly influenced by obstacles, while others appear unaffected. Distinct cases include attractive or repulsive interactions: the former is associated with velocity conservation and pronounced reorientation, whereas the latter involves less marked reorientation but a noticeable reduction in velocity.}
 
\begin{figure*}[!ht]
  \centering
  \includegraphics[width=1\linewidth]{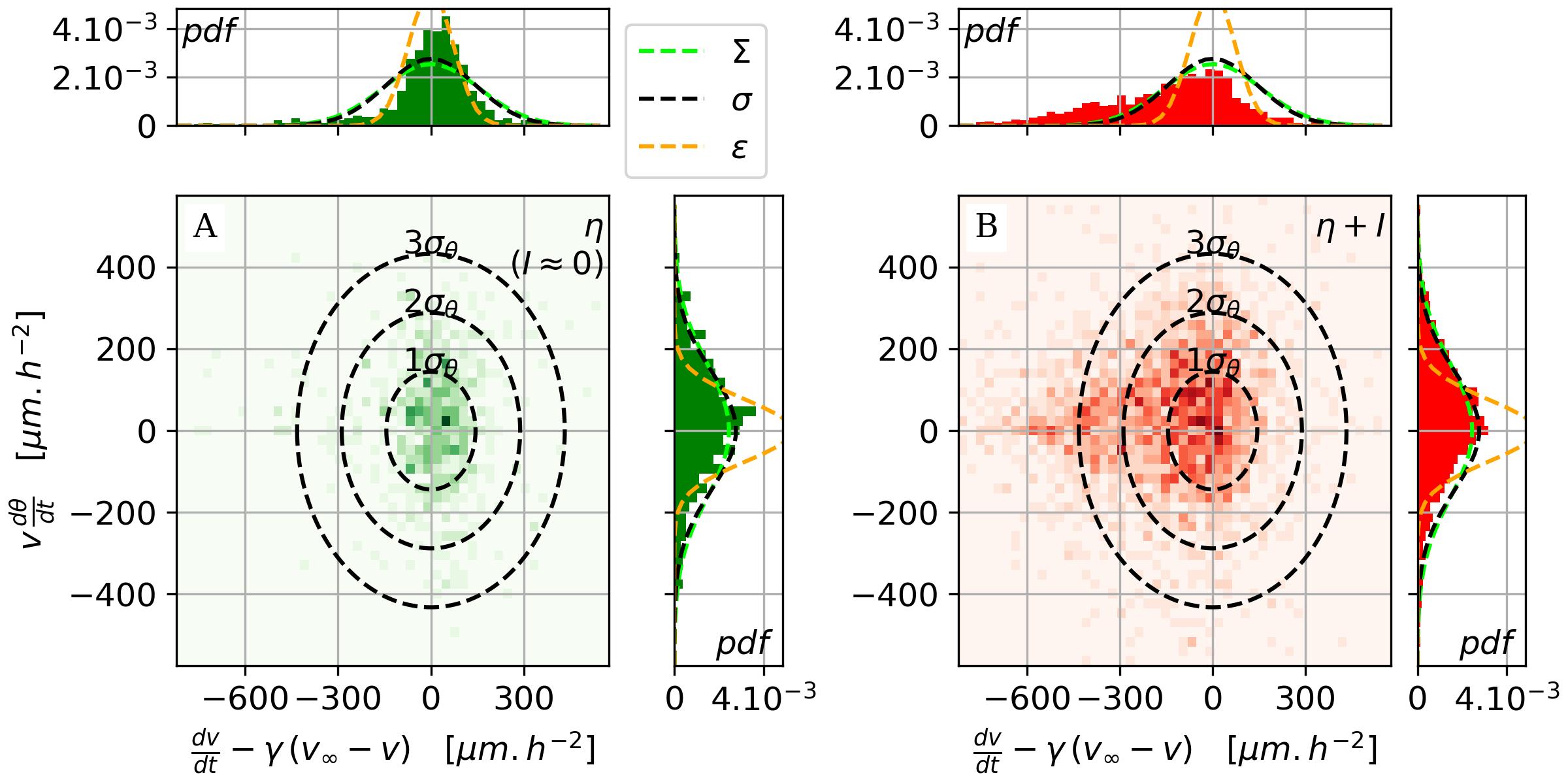}
  \caption{Example of 2D histograms $\vec{\eta}+\vec{I}$ estimations made from the model for free branches (A), where $\vec{I} = \vec{0}$, and obstructed branches (B) for {the network A}. Green dashed lines correspond to normal law centered at zero and of same standard deviation as the empirical distribution of free branches $\Sigma$ ; orange dashed lines correspond to numerical error estimations $\epsilon$ ; black dashed lines correspond to the normal law centered at zero and of standard deviation $\sigma_X = \sqrt{\Sigma_X^2-\epsilon_X^2}$ where $X$ corresponds to either $\theta$ or $v$, depending on the projection to which it refers. Dashed circles centred at $\vec{0}$ are of radius \numlist{1;2;3} $\sigma_\theta$ All theses lines are reported on the (B) side for comparison between free and obstructed branches.}
  \label{fig:noise_interaction}
\end{figure*}

We now turn to the description of $\eta$. The corresponding distributions (with expected $\vec{I}=\vec{0}$) of $\eta_v$ and $\eta_\theta$ are shown figure~\ref{fig:noise_interaction}-A. Results for the five networks are shown in table \ref{tab:tab_results}.
It corresponds to the 2D histogram of $\eta_v = \frac{\textrm{d}\vec{v}}{\textrm{d}t} - \gamma \, (v_\infty-v)$, projection of $\vec{\eta}$ in the growth direction, and of $\eta_\theta = v \, \frac{\textrm{d}\theta}{\textrm{d}t}$, projection of $\vec{\eta}$ perpendicular to the growth direction. 

$\sigma_\theta$ is estimated from the standard deviation $\Sigma_\theta$ of $\eta_\theta$ distribution corrected by the numerical error $\epsilon_\theta$: $\sigma_\theta^2 = \Sigma_\theta^2-\epsilon_\theta^2$.
Dashed circles centred at $\vec{0}$ are of radius \numlist{1;2;3} $\sigma_\theta$.
Projected distribution $\eta_v$ and $\eta_\theta$ are compared to the expected normal probability density function of mean $0$ and of standard deviation $\sigma_\theta$. 
For each thallus, distribution of $\eta_\theta$ is compatible with a normal law with standard deviation $\sigma_\theta$. 
Using Kolmogorov-Smirnov tests {do not} allow to distinguish the distribution of $\eta_\theta$ from a normal distribution with the same mean and standard deviation with a p-value of \qty{0,05}{}. 
The mean of the observed distributions (at maximum \qty{25}{\um\per\hour\squared}) is always smaller than the assumed error in the acceleration reading (of the order of \qty{56}{\um\per\hour\squared}).
See the appendix~\ref{sec:annex_temporalResolution} for details. 
The distribution of $\eta_v$ is found narrowed around 0 with more frequent rare extreme events.
This suggest{s} that $\eta_v$ does not follow a normal distribution and it is confirmed by Kolmogorov-Smirnov tests (p-value \qty{0.99}{}).
It is known~\cite{sanchez-orellana_automated_2018} that hyphae can undergo phases of acceleration interspersed with stationary phases.
This phenomena could explain the more frequent extreme events seen in the distribution of $\eta_v$.
Given the low $\sigma_\theta$ value found, which is of the same order of magnitude as the initial acceleration after branching, our model recovers and quantifies the largely rectilinear behaviour observed.

{The Langevin framework relies on stationary and Gaussian noise assumptions that are only partially satisfied here. Nevertheless, it provides a tractable and validated baseline for free trajectories. The good agreement between the apical extension speed and the model without interaction proves that the unobstructed branches do indeed appear to be undisturbed by the rest of the mycelium. 
Therefore, filtering by interaction area occupancy seems to be an effective way to select branches that interact with the rest of the thallus. }

\subsection{Interacting apical trajectories}

Let us now move on to the second collection of apical extension trajectories with suspected interactions, for which $\vec{I}\neq 0$ in equation \ref{eq:model_vect}.
{Panel B} of the figure \ref{fig:noise_interaction} show the distribution of $\vec{\eta}+\vec{I}$ for branches for which a section of thallus has been observed at least once in the interaction area.
For comparison, we superimposed the centred normal distribution with standard deviation $\sigma_\theta$ obtained in the case without interaction (dashed black lines). 

It can be clearly seen that distribution is dramatically affected, both projections $\eta_v+I_v$ and $\eta_\theta+I_\theta$ are modified. {There is no global rotation effect, the mean is zero, and there are no abrupt reorientation events in both cases.}
In the direction of growth the distribution of $\eta_v+I_v$ is almost uniformly spread out between $-5\sigma$ and $1\sigma$. 
This corresponds to deceleration events and can be interpreted as the consequence of rapid, large-scale reorientation. 
The distribution in the direction perpendicular to the growth of $\eta_\theta+I_\theta$ is centred and symmetrical. This indicates that the reorientations are not biased in one direction (no chirality) and that they are brief, typically of the order of one time step. 
The distribution also appears to be narrower at zero. 
This suggest{s} that, after an eventual initial change of direction, the growth direction is more constrained to the straight line by the interactions than it was in the case of unobstructed growth.

{The interaction term $I$ breaks the Gaussian assumption underlying the standard Langevin framework. The deviations observed for obstructed trajectories are sufficiently pronounced to be robust to these approximations. This confirms that interactions cannot be captured by a simple renormalization of the noise amplitude $\eta$, but require an additional non-Gaussian term $I$ whose precise mathematical expression — likely dependent on the local geometry and density of surrounding hyphae — goes beyond the scope of the present work. The Langevin framework nonetheless provides a useful baseline: by characterizing free trajectories rigorously, it allows interaction events to be identified as statistically significant deviations from Gaussian dynamics.}

\section{Discussion and Conclusion}
\label{sec:discussion}

We propose a modelling of the trajectories of hyphal extension based on the well-known advection-diffusion equation, that combines a growing force $\vec{F_g}$, a friction force $\gamma \, \vec{v}$, a noise of biological origin $\vec{\eta}$ and a force linked to self-interaction $\vec{I}$.
This model satisfactorily reproduces the trajectories of freely moving apexes, notably capturing the asymptotic velocity $v_\infty$, the initial acceleration $\gamma$, and the intensity of biological noise $\sigma_\theta$.
The random force linked to the variation in the position of the SPK, $\vec{\eta}$, is about four times weaker than the growth force ($\gamma \,  v_\infty/\sigma_\theta\approx 4$), leading to observed trajectories that are largely rectilinear, with no chirality. 

The apical growth pattern observed in \textit{P. anserina} is rapid and isotropic, driven by a biological noise that is sufficiently large to enable homogeneous densification, yet small enough to permit exploration over extended spatial ranges. Such a network architecture is particularly suited for foraging, facilitating the rapid capture, exploitation, and defense of newly available territory~\cite{simonin_physiological_2012}, especially within highly competitive environments such as herbivore feces.

Decomposer fungi continually exhaust the organic substrates on which they grow and feed, and therefore depend on the repeated successful colonization of new resources. In terrestrial ecosystems, such resources are heterogeneously distributed across space and time. As proposed in \cite{boddy_fungal_2016}, colonization success can be decomposed into two components: (i) the ability to arrive at, penetrate, and establish within a resource, and (ii) the ability to persist within that resource until reproduction and subsequent dissemination. Consequently, the processes of arrival and spread are critical to the ecological success of saprotrophic fungi. Following resource arrival, competitive ability determines the extent of colonization and the duration of territorial retention.

We assumed that biological noise is isotropic; however, experimental observations reveal that only the fluctuations of $\eta_\theta$ follow a Gaussian distribution centred at zero. In contrast, the distribution of $\eta_v$ is both more sharply peaked—due to the combined effects of numerical and biological noise around $v_\infty$—and exhibits heavy tails, reflecting the presence of extreme events associated with stop-and-go dynamics. \\
It then becomes possible to detect an effect of self-interaction. 
We find that this interaction leads to a statistically significant and pronounced reorientation of apical trajectories.
The effect is both intense and short-lived, with the most prominent manifestation being a marked deceleration at the moment of reorientation.

{
Analyzing this interaction is delicate and lies beyond the scope of the present work. We can, however, outline a set of questions that illustrates the complexity of the problem. Obstacles are defined by their presence within a disk sector. Beyond this spatial proximity, interactions may depend on memory effects (path dependence), on dynamical effects (magnitude and orientation of velocity), on the absolute age of the thallus or apex, or on the relative age of the apex with respect to the obstacle. Environmental dependence may also be considered, for instance a differential effect in nutrient-rich versus nutrient-poor media. A key challenge would be to predict, given a set of initial conditions, whether an interaction will be attractive or repulsive. Finally, since apical trajectories are inherently noisy, distinguishing an individually impacted trajectory from one influenced by an obstacle is non-trivial.}

To analyse these interactions in detail, they must be distinguished first.
A first approach would be to reconstruct the histogram in figure~\ref{fig:noise_interaction}-B as a mixture of free and constrained behaviours, thereby extracting interaction events. 
This is challenging because such interactions, sometimes contradictory, can occur simultaneously in a dynamic environment.
As shown in figure~\ref{fig:ExampleInteractions} and the accompanying movie, the apical hypha whose dynamics are being tracked both grows and branches, while the hyphae forming the obstacle also grow and branch.
One way to address this complexity is to isolate a single hypha, which can be achieved in a microfluidic setup~\cite{baranger_microfluidic_nodate}. 
However, this configuration departs substantially from natural growth conditions, and the full spectrum of interactions is difficult to reproduce.

Evidence of self-interaction implies a form of long-range communication enabling self-recognition during vegetative growth.
Such an effect has also been observed in other species \cite{simonin_physiological_2012}. 
One explanation could be a chemotropism sensitive to a molecule produced by the apexes during their growth. 
In \cite{Clark-Cotton2022Feb} a modelling was proposed that takes into account diffusion of chemical species into the growing media to guide growth. 
{Chemotropism is a well-documented mechanism in many organisms, where gradients of signaling molecules—such as nutrients, hormones, or secondary metabolites—direct growth orientation with high spatial precision. Such gradients can provide robust positional information even in heterogeneous environments, supporting the plausibility of a chemically mediated self-recognition process~\cite{turra_hyphal_2016,sadaf_chemotrophic_2024}.}
To test this hypothesis, it would be necessary to observe the correlation between reorientation, identification of the obstacle across the interaction area and the time elapsed since the obstacle grew at this point.

\printbibliography
\clearpage
\onecolumn
\appendix
\clearpage
\section{Supplementary Material}
\subsection{Data accessibility statement }
Data and code are accessible through a \href{https://github.com/thibault-chassereau/HyphalNetwork_ApexHypha_Interactions}{github} repository. 
Code is written in usual Pyhton-3. A JupyterNotebook is proposed that merges all the numerical work.

\subsection{Discretisation and estimation of growth elongation vectors}
If we define the point $A_i$ to be the $i$-th apex of a branch, at time $t_i$, at position $(x_i,y_i)$ and corresponding to curvilinear abscissa $s_i$ then we can define at the time $\tilde{t}_i = \frac{t_i+t_{i+1}}{2}$ the following:

\begin{equation}
    \left\{ 
    \begin{array}{ccl}
        v_i & = &\frac{s_{i+1}-s_{i}}{t_{i+1}-t_{i}}\\
        \theta_i & = &\arctan{\left(\frac{y_{i+1}-y_i}{x_{i+1}-x_i}\right)}\\
        \frac{dv}{dt}_i & = & \frac{v_{i+1}-v_{i-1}}{(t_{i+2}+t_{i+1})/2-(t_i+t_{i-1})/2}\\
        \frac{d\theta}{dt}_i & = & \frac{\theta_{i+1}-\theta_{i-1}}{(t_{i+2}+t_{i+1})/2-(t_i+t_{i-1})/2}
    \end{array} 
    \right.
    \label{eq:discretisation}
\end{equation}
 where $v$ corresponds to the norm of the growth vector, $\theta$ to its orientation, $\frac{dv}{dt}$ the change in the norm of the growth vector and $\frac{d\theta}{dt}$ the change in its orientation.

\subsection{Propagation of measurement error}

This section deals with the propagation of the error on the position of the apexes on the estimation of the $\eta_v$ and $\eta_\theta$ accelerations studied in the article.
We assume that the error in position is the same in both directions of the plane and we denote it $\delta x$.
Since the apparent diameter of hyphae is around \SI{10}{\um}, we estimate this uncertainty at a quarter of a hyphal diameter, i.e. \SI{3}{\um}.
Considering that the two directions are independent, it follows that the error on $\eta_v$, $\delta\eta_v$ is equal to:
\begin{equation}
    \delta \eta_v = \frac{2\delta x}{T^2}\sqrt{1+\frac{\gamma^2T^2}{2}}
\end{equation}
and the error on $\eta_\theta$ is equal to:
\begin{equation}
    \delta\eta_\theta = \frac{2\delta x}{T^2}\sqrt{1+\frac{\eta_\theta^2T^2}{2v^2}}
\end{equation}
where $T=\SI{18}{\min}$ is time between 2 timesteps.
In order to get a numerical estimation, we consider the case where $\eta_\theta\approx\sigma\approx\SI{150}{\um\per\hour\squared}$ and $v\approx v_\infty\approx \SI{300}{\um\per\hour}$. 
We then find $\eta_\theta\approx \SI{67}{\um\per\hour\squared}$.

\subsection{Time resolution and numerical error.}\label{sec:annex_temporalResolution}

If the delay between two images is too short then the error in the measurement will be too big to extract any information about the biological noise. 
At the opposite, if this delay is too large then we can only see an average version of what is really happening during the growth and we can miss on little variation.

To help us find a good compromise, we used a mycelial network of \textit{P. anserina} that had grown under the same standard conditions but for which we had a time resolution of \qty{4}{\minute} between each image.
By subsampling this network we can measure the effect of the time resolution on the biological noise estimation.

\begin{figure}[!ht]
    \centering
    \includegraphics[width=\linewidth]{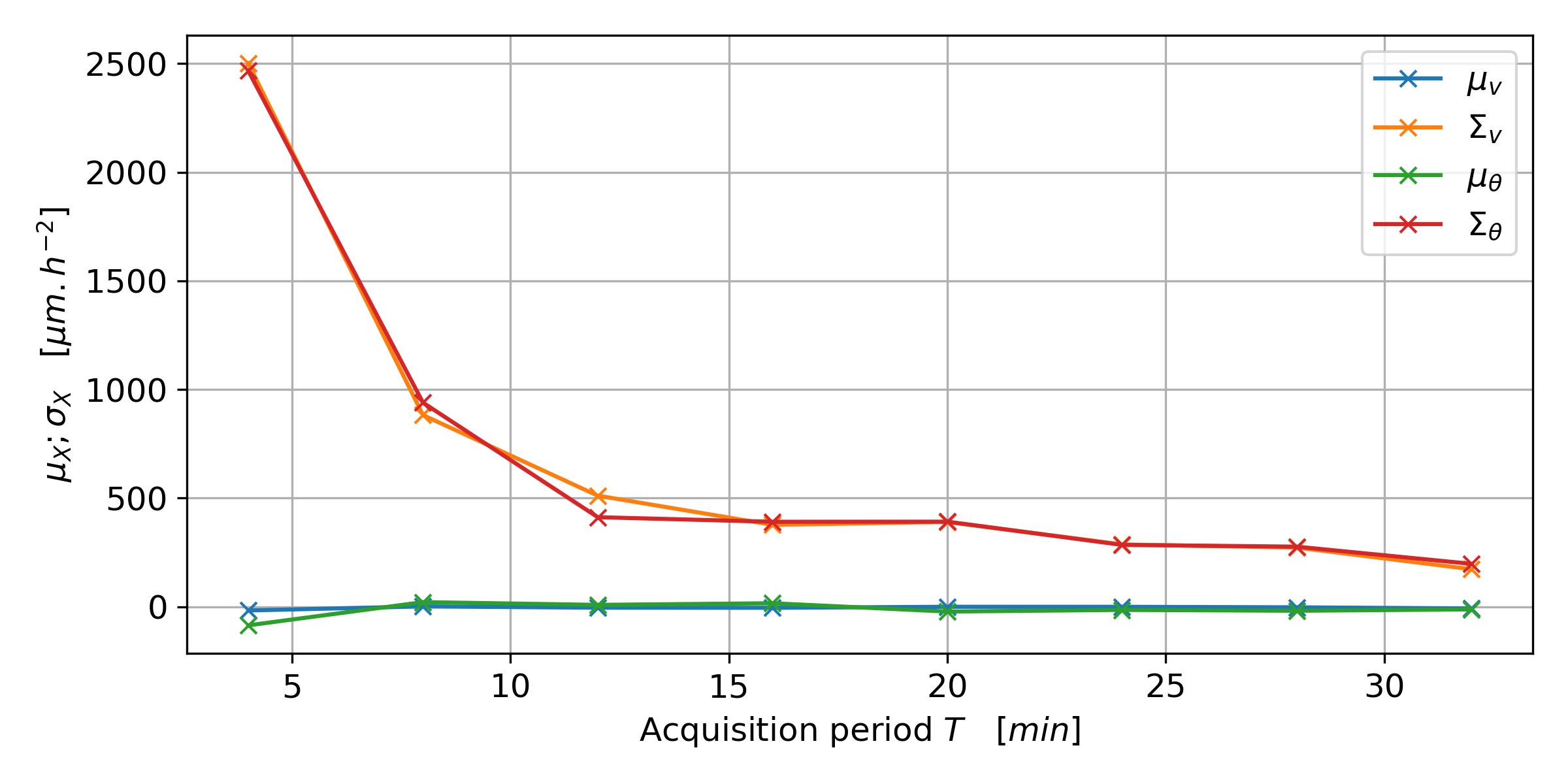}
    \caption{Characteristic (means $\mu$ and standard deviations $\Sigma$) of the distributions of $\eta_v$ and $\eta_\theta$ depending on the delay $T$ between two consecutive images.}
    \label{fig:subsamplingTimeResolution}
\end{figure}

In figure \ref{fig:subsamplingTimeResolution}, we show the evolution of the characteristics (means $\mu$ and standard deviations $\Sigma$) of the empirical distributions of $\eta_v$ and $\eta_\theta$ as functions of the delay between consecutive images.

As predicted, a short time period between image leads to very large distributions as the numerical error is maximized.
When the delay between images increase, the distributions converge toward the distribution of biological noise only as the numerical error decrease.
After a 20-minute delay between two images, the benefits in terms of reducing digital noise are outweighed by the difficulty of reconstructing the growth of the network with any certainty.
A good compromise was therefore found by taking a delay of 18 minutes between each image.

\end{document}